%% file: main.tex
\documentclass[10pt,conference]{IEEEtran}

\AtBeginDocument{%
  \providecommand\BibTeX{{%
    Bib\TeX}}}

\usepackage{subfiles}
\usepackage{graphicx}
\usepackage{listings}
\usepackage{color}
\usepackage{graphicx}
\usepackage{subcaption}
\usepackage{footmisc}
\usepackage{tcolorbox}
\usepackage{xr}
\usepackage[hyphens]{url}
\usepackage{hyperref}

\usepackage{cite}
\usepackage{amsmath,amssymb,amsfonts}
\usepackage{algorithmic}
\usepackage{graphicx}
\usepackage{textcomp}
\usepackage{xcolor}
\def\BibTeX{{\rm B\kern-.05em{\sc i\kern-.025em b}\kern-.08em
    T\kern-.1667em\lower.7ex\hbox{E}\kern-.125emX}}

\definecolor{dkgreen}{rgb}{0,0.6,0}
\definecolor{gray}{rgb}{0.5,0.5,0.5}
\definecolor{mauve}{rgb}{0.58,0,0.82}
\definecolor{darkred}{rgb}{0.6, 0, 0}

\def\tool{RefExpo}
\def\datasetCount{20}

\graphicspath{ {./images/} }

\begin{document}

\title{\tool: Unveiling Software Project Structures through Advanced Dependency Graph Extraction}

\author{
    \IEEEauthorblockN{Vahid Haratian\IEEEauthorrefmark{1}, Pouria Derakhshanfar\IEEEauthorrefmark{2}, 
    Vladimir Kovalenko\IEEEauthorrefmark{2}, Eray Tüzün\IEEEauthorrefmark{1}}
    \IEEEauthorblockA{\IEEEauthorrefmark{1}Bilkent University, Ankara, Turkey \\
    Email: \{vahid.haratian, eraytuzun\}@bilkent.edu.tr}
    \IEEEauthorblockA{\IEEEauthorrefmark{2}JetBrains Research, Amsterdam, The Netherlands \\
    Email: \{pouria.derakhshanfar, vladimir.kovalenko\}@jetbrains.com}
}

\maketitle

\begin{abstract}
\subfile{sections/abstract}
\end{abstract}



\begin{IEEEkeywords}
Software Analytics, Dependency Graph, Call Graph, Code Referencing, Software Structure, Software Architecture
\end{IEEEkeywords}


\maketitle

\section{Introduction}
\subfile{sections/introduction}

\section{\tool{}}
\subfile{sections/tool}

\section{Implementation}

\subfile{sections/implementation}

\section{Dataset}
\subfile{sections/dataset}

\label{sec:dataset}

\section{Evaluation}
\subfile{sections/evaluation}

\section{Limitations}
\subfile{sections/limitations}

\section{Conclusion}
\subfile{sections/conclusion}

\section* {Acknowledgements}
\subfile{sections/acknowledgements}

\bibliographystyle{IEEEtran}
\bibliography{references}

\end{document}

%% file: sections/abstract.tex
The dependency graph (DG) of a software project offers valuable insights for identifying its key components and, hence has been leveraged in numerous studies. 
Nevertheless, there is a lack of reusable tools for DG extraction.
Existing tools are either outdated and difficult to configure, or fail to provide accurate analysis.
However, Integrated Development Environments (IDEs) are designed to address the above issues.

This study introduces \tool{}\footref{tool}, a reusable DG extraction tool that supports multiple languages, such as Java, Python, and JavaScript.
\tool\ is a plugin based on IntelliJ which is a well-maintained and reputed IDE.
In addition, we compile an initial version of our dataset consisting of \datasetCount\ Java and Python projects.

We evaluated \tool{}'s validity at two levels: specific language features and comparisons against other existing tools, which we refer to as the micro and macro levels.
Our evaluation shows \tool\ achieving 92\% and 100\% recall on micro test suites Judge and PyCG for Python and Java, respectively.
In macro-level experiments, \tool\ outperformed existing tools by at least 31\% and 7\% in finding unique and shared results (non-overlapping and overlapping with other tools).

The installable version of \tool\ is available on the IntelliJ marketplace\footref{market}.
Additionally, a short video describing its functionality is available on YouTube\footnote{\url{https://youtu.be/eCnPUlj6YgA}}.

%% file: sections/introduction.tex
Numerous studies have focused on analyzing the dependency graph (DG) of software projects~\cite{bh-swGraph, ma-serviceDependency, deng-systemDependence, cai-programDependence, jian-ooDependencies, walunj-callGraph, ryder-callGraph, pycg, reif-opal, Keshani2023}.
A dependency graph typically represents software components as nodes and their logical dependencies as edges.
In such a graph, nodes can be methods, classes, files, modules, and etc. Logical dependencies are identified by how each component references others.
These references can be method calls, property access, inheritance relationships, and etc.

In addition to studies that solely focus on analyzing DGs, there are studies primarily aimed at utilizing DGs to extract insights about their significance~\cite{inoue-sigRanking, wang-keyClass, pan-elementRank, li-keyClass}, comprehensibility~\cite{sora-comprehension, kamran-comprehension, yang-condensing, thung-condensing}, and influence of each software project component~\cite{tie-changeImpact, li-influence, gu-influence}.
Moreover, many software analytics studies utilize DG insights to focus their assessments on the most critical components of a project~\cite{zimmermann-subsytemFailures, zimmermann-defectPrediction, nguyen-predicting, du-bugPrediction, meilong-defectPrediction}.
However, despite the abundance of such studies, there is a notable lack of reusable tools for generating dependency graphs~\cite{pycg}.
This gap highlights the technical challenges researchers face in extracting such data, often restricting them to narrow the scope of their studies to small projects~\cite{pycg}.

Although there are many open-source and commercial tools for creating dependency graphs, they are often difficult to use.
Some of these tools are no longer maintained, making them outdated and unable to support new language features, which leads to syntax issues~\cite{pyan, pycg, pyne}.
For example, we observed that the majority of Python tools such as PyCG~\cite{pycg} and Pyan~\cite{pyan} are not compatible with the current version of Python (3.12 at the time).
Therefore, they could not analyze the actively maintained popular GitHub repositories.
Nevertheless, we tried using older versions of the repositories, but the tools were highly sensitive to the Python version, and we could only find one usable revision in an experimental repository.

Besides, some of the tools function adequately only for small projects and are difficult to configure for larger ones~\cite{jarviz, sonargraph, df, ndepend}.
Although the definition of a small-sized project can vary, we have found it challenging to use these tools on some popular GitHub repositories.
For example, the majority of Java tools provide their analysis based on a built JAR or WAR file.
However, a big project like ElasticSearch does not compile into a simple WAR file that these tools can process.
Besides, this approach excludes the code that will not be included in the final built artifacts, such as test cases, build scripts, and documents.

On another track, tools that provide evaluation over source code presented significant challenges in their configuration
For instance, Sonargraph~\cite{sonargraph} requires manual configuration over the build tools to be able to provide its evaluations.
This can be done easily on small projects; however, for a larger project with multiple levels of submodules, doing so is not an easy task.
For example, the complex build structure of ElasticSearch makes it difficult to configure Sonargraph probes, and we were unable to do so in our attempts.

Among those tools that evaluate the source code directly, there are tools that do not require much configuration.
However, their accuracy is so low that it is inadequate for the majority of use cases~\cite{rexdep}.
These tools mainly rely on explicit imports at the beginning of the files to capture references.
However, they can easily fail to find the target of the import due to the complexity of the project structure.
Moreover, this excludes a significant portion of references that occur within packages and files.

Therefore, to fill this gap and facilitate dependency extraction, in this study, we introduce \tool{}~\footnote{\url{https://github.com/vharatian/RefExpo}\label{tool}}. 
It is a reusable tool that we utilized in our previous study BFSig~\cite{bfsig} to generate DG.
\tool\ is based on IntelliJ IDEA, one of the most robust Integrated Development Environments (IDEs) available.
IDEs like IntelliJ need to have a sophisticated dependency analysis to offer robust code navigation and reference identification.
Since IDEs are designed for compatibility and they work with the compiler provided to them, they stay updated with the latest language versions while supporting the older ones.
Moreover, their easy-to-use interface makes it easy and straightforward to import and manage projects compared to the experimental tools.
Although the importing phase is easy and automatic for most projects, large OSS projects with complicated structures usually provide guidance on how to configure IDEs to gain full code recognition.
For instance, ElasticSearch, and Apache Flink, provide documentation on how to import their projects into IntelliJ.

Finally, we believe that the unavailability of reliable datasets might compromise the validity of some studies due to the difficulty of data extraction.
Some studies can be so sensitive to minor changes that they significantly impact study outcomes~\cite{du-bugPrediction}.
This challenge inspired us to create a dataset to potentially accelerate studies that require DGs.
This allows researchers who want to include DGs in their projects to concentrate on their studies without the burden of DG extraction.
We have compiled an initial dataset~\footref{tool} of \datasetCount\ popular GitHub projects, encompassing Java and Python, with an eye toward expansion.

%% file: sections/tool.tex
\tool\ generates comprehensive dependency graphs from projects developed in a variety of programming languages. 
Utilizing IntelliJ as the underlying engine, \tool\ can support projects consisting of multiple programming languages including Java, Python, JavaScript, Go, Ruby, and HTML, and also framework-specific syntax.
For instance, it is possible to add Python support by adding a Python plugin, which is provided by the internal JetBrains team.
In addition, by adding framework support (for instance, Angular), \tool\ can provide a more sophisticated dependency graph for custom framework implementations.
Finally, the utilization of an advanced engine such as IntelliJ allows \tool\ to evaluate large-scale projects with zero to no required configurations.
Nevertheless, in large projects with custom structures, some configurations might be required.
However, due to the popularity of IntelliJ, these projects usually share a guide on how to configure the project for this IDE.

The tool comes in the format of a plugin that can be installed on IntelliJ through its marketplace\footnote{\url{https://plugins.jetbrains.com/plugin/23684-refexpo}, \label{market}} or could be used with IntelliJ headless mode in the command line through the source code\footref{tool}. 
To use \tool{}, the user needs to import their project into IntelliJ and ensure that IntelliJ is properly configured to index the source code, and then open the \tool\ toolbox.
No project build is required, but to make sure that the project is imported correctly and is indexed by IntelliJ, one can try to build the project before running the evaluation.
To manage large projects effectively and enhance usability, \tool\ offers configurable filters.
The filters allow users to apply regular expressions for evaluating files, classes, and methods. 
Additionally, \tool\ detects circular references and provides options to filter them out at the file, class, and method levels.
While these filters could be applied to the CSV file after the evaluation, applying a filter beforehand could help avoid generating an excessively large output file.
Besides, evaluating all files with all references in a large project might result in long execution which can be sped up by prior filtering.

Finally, the tool produces a CSV file that contains all identified references. 
The graph is represented as its edges in the CSV file by providing a source and target locator. The locators are identified with three main types of attributes, File, Class, and Method. 
Table~\ref{tab:output} shows the information provided for each locator in the CSV file. 
In parts of the literature~\cite{ryder-callGraph, walunj-callGraph}, graphs that include dynamic links are referred to as call graphs, while those containing only static references are called dependency graphs. 
We will continue using these terms accordingly. 
You can refer to a complete usage guide including a short video, available in the replication package\footref{tool}.

\begin{table}[]
\resizebox{\columnwidth}{!}{%
\begin{tabular}{cl}
\hline
\begin{tabular}[c]{@{}c@{}}Parameter Name \\ (source, target)\end{tabular} & \multicolumn{1}{c}{Description}                                                                                                                     \\ \hline
Path                                                                      & File path that contains the location                                                                                                                \\
Line                                                                      & Line number in the file                                                                                                                             \\
Class                                                                     & Class name that contains the location                                                                                                               \\
ClassFull                                                                 & Class name including the package name                                                                                                               \\
Method                                                                    & Method name that contains the location                                                                                                              \\
MethodFull                                                                & Method name including the class and package                                                                                                         \\
Structure                                                                 & \begin{tabular}[c]{@{}l@{}}Nested structure of the location in file like method \\ defined inside method or class defined inside class\end{tabular} \\ \hline
\end{tabular}%
}
\caption{Locator Attributes in \tool\'s CSV output }
\label{tab:output}
\end{table}


%% file: sections/implementation.tex
To understand the process and intricacies of \tool{}'s inspection mechanism, it is essential to delve into its two-phase strategy.

\begin{figure}
  \includegraphics[width=\columnwidth]{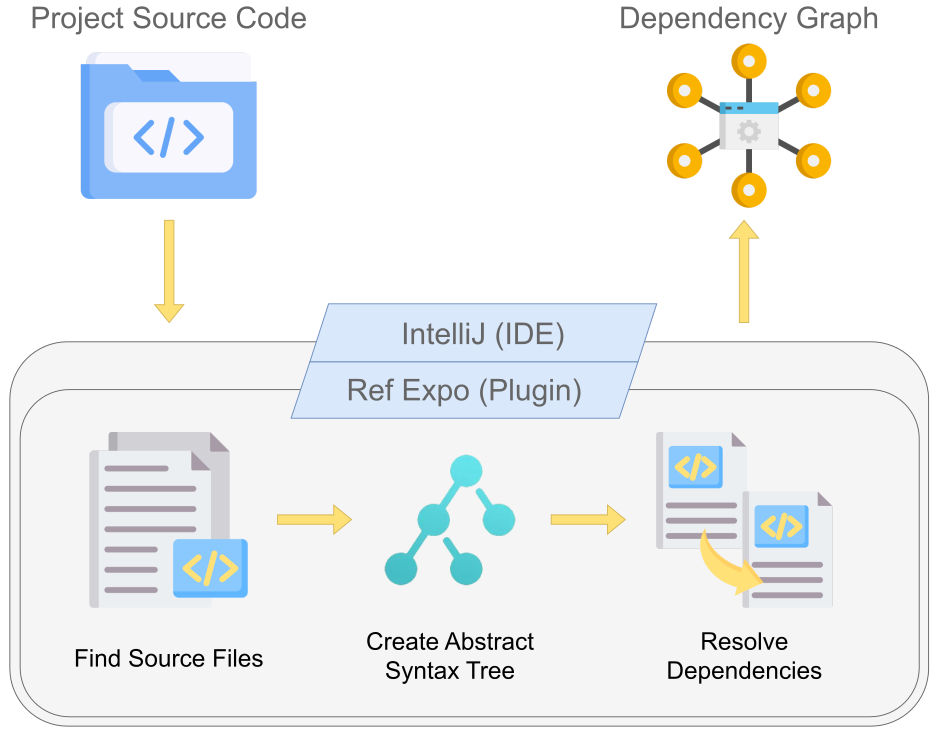}
  \caption{\tool\ general structure}
  \label{fig:structure}
\end{figure}

\tool\ has two inspection phases that are quite similar to each other.
Due to the lazy indexing policy of IntelliJ, \tool\ needs to administer a two-pass strategy.
Lazy indexing in IntelliJ postpones the indexing process until a file is opened and indexing is needed, rather than keeping the user waiting for the entire indexing to complete.
This is in contrast to what \tool\ requires, which is the complete indexing before running the evaluations.
Therefore, in the first path, \tool\ will traverse all possible files and do a full traverse over their Abstract Syntax Tree (AST), which is provided by the PSI API of IntelliJ plugin SDK~\cite{intelliJ-SDK}.
For each element in the AST, IntelliJ provides a list of places in which the statement references.
In the first path, \tool\ only resolves all the references, which makes the IntelliJ index target files of the reference and will keep this information.
Afterward, \tool\ does another traversing round, which will go over all files and elements again. 
However, this time, it will evaluate each resolved reference and log its encounter.
Figure \ref{fig:structure} shows the general structure of \tool{} inspection.

\tool\ focuses on internal referencing and does not evaluate external references.
For instance, it is usual to reference \texttt{java.lang.String} in a Java project, however, \tool\ filters out all those references.
To achieve this, \tool\ initially filters out project files to those included in the version control system (VCS) of the project.
This will also exclude the build products in addition to the dependency files since IntelliJ is capable of reading and providing references in already compiled CLASS and JAR files.
Finally, the target of each reference should be included in the VCS so \tool\ consider it as a valid reference in the project. 

Ultimately, \tool\ generates a CSV file listing all encountered references. 
Each reference will include three locators -- File, Class, and Method -- for both the source and target. 
Depending on the reference type, it may have one, two, or all three of these locators.
For instance, since HTML files do not contain classes or methods, a reference to or from an HTML file will be labeled only by the file name.
Consequently, the class and method columns will be empty in the CSV file.

%% file: sections/dataset.tex
To help researchers quickly access an extracted dependency graph, we have evaluated \datasetCount\ repositories developed in Python and Java~\footref{tool}.
Our selection process involves identifying popular projects on GitHub and utilizing its search API. 
We begin by ranking repositories according to their star count and subsequently gather data on their programming languages. 
Following this, we specifically filter for repositories primarily developed in Java and Python.
We manually evaluated and removed repositories that are small code samples from the resulting repositories.
Finally, we have added those projects that fall out of this filtering process but we have evaluated them during our macro evaluation phase.
Those projects were initially selected due to their compatibility with other tools which we compared \tool\ results with.
As it has been presented in our replication package~\footref{tool}, the dataset contains projects with different sizes from only 11 files to 15K.
This approach yields a comprehensive dataset designed to assist researchers with diverse requirements.
Our objective is to continually expand this dataset, and we are open to collaborations with researchers, offering to tailor the dataset to meet their specific needs.

%% file: sections/evaluation.tex
In this section, we provide an overview of our evaluation methods and results.
A detailed description of our evaluation methods, metrics, a complete results table, and accompanying diagrams can be found in Appendix C\footref{appendix}.

We evaluated \tool{}'s accuracy and effectiveness in identifying dependencies at two different levels micro and macro.
On the micro-level, we evaluate how well \tool\ captures specific features of the target languages. 
On the macro-level we evaluate the performance of \tool\ against existing tools over real-world projects.

\subsection{Micro Level}
We utilized two test suites, Judge and PyCG\cite{reif-judge,pycg}, which provide test cases for all the features of Java 8 and Python 3.7.
These test suites are originally designed for call graphs.
There is a logical difference in the definition of call graphs and dependency graphs. 
Call graphs are only concerned about the possible paths and routes that control flow might navigate during execution as a result of function invocation~\cite{ryder-callGraph, ferrante-flowdependency}
In contrast, dependency graphs account for all possible components referenced, where method call is one of them.
For example, a simple property access or inheritance relationship is considered as a reference in a DG, however, it will not necessarily be an edge in a call graph, as it might not alter the program's control flow. 

Since our focus is on dependency graphs, while we evaluated \tool\ using the original test suites, we adapted them to conduct a more relevant round of evaluation.
To adapt the test suites, we removed test cases involving dynamic referencing and external links.
You can read a full discussion about the differences and examples of both graphs, in addition to our dataset cleanup process in Appendix B\footref{appendix}.

We used recall as the performance indicator to determine how much of the specified results are captured by \tool{}. 
We found accuracy irrelevant in this context since \tool\ is designed for DGs and, therefore, produces many results that are not included in this dataset.

You can find the full evaluation results separated by features in Appendix C\footref{appendix}. 
However, our results indicate that:

\begin{tcolorbox}[colback=gray!10!white, colframe=gray!75!black, title=Micro Evaluation Results]
    \tool\ achieved 100\% recall on the cleaned version of the Judge Test Suite, and 85\% on the initial suite. 
    For the PyCG Test Suite, it achieved 97\% recall on the cleaned version and 68\% on the initial suite.
\end{tcolorbox}

\subsection{Macro Level}
We compared the outputs of \tool{}'s against five popular tools in eight popular GitHub repositories. 
We assessed \tool\ against Jarvix, Dependency Finder, and Sonargraph for Java, and Pyan and PyCG for Python.

Since there is no ground truth to compare each tool's results against, we compared the tool results against each other.
On one hand, we evaluate the overlap of the outputs of different tools as a measure of the tool's agreement with others.
These overlaps can be only between two of the tools or more (all).
On the other hand, we evaluate the unique results each tool reports as a measure of the uniqueness of each tool's output in identifying edges that other tools cannot find.

We calculate the unique ratio as the ratio of unique results to the total number of edges.
Similarly, we calculate the shared ratio as the count of the edges that have been identified by each tool and at least one other tool, relative to the total number of edges.
It should be noted that all these ratios are calculated based on the total number of edges that have been identified by all the tools.
Hence, for each tool, the ratio of unique edges plus the ratio of shared edges does not equal one but represents the ratio of its reported output to the total edges.

You can find the full comparison of the outputs including visualizations separated for each project in Appendix C\footref{appendix}. 
However, our results indicate that:

\begin{tcolorbox}[colback=gray!10!white, colframe=gray!75!black, title=Macro Evaluation Results]
For Python projects, \tool\ covered 90\% of identified edges, outperforming Pyan (33\%) and PyCG (26\%). 
\tool\ had 35\% shared and 56\% unique edges. 
For Java projects, \tool\ covered 92\% of identified edges, surpassing Jarviz (41\%), Dependency Finder (29\%), and Sonargraph (55\%). 
\tool\ had 59\% shared and 34\% unique edges.

\end{tcolorbox}

%% file: sections/limitations.tex
\tool\, like other tools, has some limitations.
When it comes to tracing dynamic references between components, \tool\ is not the best compared to other available dependency extraction tools. 
\tool\ is optimized to evaluate dependency graphs and is not designed to provide the most accurate results for call graphs.
Nevertheless, our evaluations show that \tool\ can yield a good enough recall and capture a significant portion of call graphs.
A detailed discussion of the differences between these graphs is included in Appendix B\footnote{\url{https://figshare.com/articles/journal_contribution/Appendix_Performance_Evalaution_of_RefExpo/26110708}, \label{appendix}}.

Moreover, although IntelliJ is capable of parsing many programming languages and \tool\ should, in theory, be able to evaluate those languages, we have only optimized and evaluated \tool's performance for Python and Java.
We have utilized our micro-evaluation process to improve \tool's performance by leveraging different IntelliJ APIs.
To provide the best results for other programming languages, we would need to repeat this optimization and evaluation cycle for each new language.

Finally, our data set has just been initiated and consists \datasetCount\ projects for Python and Java.
However, we plan to expand the dataset by including more projects and supporting additional languages such as JavaScript, and C\#.

%% file: sections/conclusion.tex
Dependency graphs contain valuable information that can be incorporated into software analytics studies.
However, existing tools are often challenging to use and limited in their support for various programming languages and artifacts.

Nevertheless, Integrated Development Environments (IDEs) must be capable of accurately and efficiently generating dependency graphs to support features like robust code navigation and precise reference identification.
Given this requirement, we built \tool\ on IntelliJ IDEA, one of the most reputable and reliable IDEs available.
IntelliJ is a well-documented and community-supported tool that ensures compatibility with modern programming languages while supporting older versions.
This integration enables \tool\ to support a diverse array of programming languages, ranging from Java to HTML, and provides a reusable interface that simplifies dependency graph generation.

In this study, we evaluated the capabilities of \tool\ for Python and Java.
Our evaluations suggest that \tool\ achieves 92\% and 100\% recall on micro test suites Judge and PyCG for Python and Java, respectively.
In macro-level experiments, \tool\ outperformed existing tools by at least 31\% and 7\% in finding unique and shared results.

Additionally, we have commenced the creation of a dataset encompassing dependency graphs from \datasetCount\ popular GitHub repositories, consisting of Java and Python projects.
This dataset aims to facilitate research in software analytics by providing ready-to-use, high-quality dependency graphs, thereby reducing the burden of data extraction for researchers.

%% file: sections/acknowledgements.tex
This study was mainly supported by the Scientific and Technological Research Council of Turkey (TUBITAK) 3501 program (Project Number: 121E584) and partially supported by JetBrains Research. 
We have received approval for a grant from the JetBrains Research team to further develop the tool and expand its support to more languages.
Special thanks to JetBrains' IntelliJ team for helping maximize the tool's potential.